\documentclass[prl,twocolumn,aps,amsmath,amssymb]{revtex4-1}

\usepackage[dvips]{epsfig}
\usepackage{float}
\usepackage{amsmath}
\usepackage{amssymb}
\usepackage{amsfonts}
\usepackage{euscript}
\usepackage{enumerate}
\usepackage{hhline}
\usepackage{threeparttable}
\usepackage{multirow}
\usepackage{supertabular}
\usepackage{pslatex}
\usepackage{tabularx}
\usepackage{color}

\usepackage{graphicx}
\usepackage{dcolumn}
\usepackage{bm}

\makeatletter

\begin{document}

\title{On-chip cavity  quantum phonodynamics with an acceptor qubit in silicon}

\author{Rusko Ruskov and Charles Tahan}

\affiliation{Laboratory for Physical Sciences, 8050 Greenmead Dr., College Park, MD 20740}

\email{charlie@tahan.com, ruskovr@gmail.com}

\begin{abstract}

We describe a chip-based, solid-state analogue of cavity-QED
utilizing acoustic phonons instead of photons.
We show how long-lived and tunable
acceptor
impurity states in silicon nanomechanical
cavities can play the role of a matter non-linearity for coherent phonons
just as, e.g., the Josephson qubit plays in circuit-QED.
Both strong coupling (number of 
Rabi oscillations $\lesssim$  100) and
strong dispersive coupling (0.1-2 MHz) regimes can be reached in cavities in the 1-20 GHz range,
enabling the control of single phonons, phonon-phonon interactions, 
dispersive phonon readout of the acceptor qubit, and compatibility with
other 
optomechanical components such as phonon-photon translators.
We predict explicit experimental  signatures  of
the acceptor-cavity system.
\end{abstract}

\maketitle

Circuit-QED has revolutionized the field of cavity-QED (cQED) \cite{Kimble:1998,Raimond:2001,Blais:2004}
providing a stable platform for light-matter interaction in the microwave regime along
with  large couplings
and solid state integrability.
Progress in the field has enabled
applications such as
single microwave photon sources \cite{Houck:2007-SinglePhoton}
and quantum logic gates \cite{Blais:2004}
on a chip.
In an ideal crystal environment,  phonons  may  play a role analogous to photons,
though they propagate with the much slower speed of sound.
That  acoustic phonons can be quantum coherent
has been explored  in a number of  architectures,
allowing seminal experiments in
optomechanical cooling
\cite{Kippenberg:2008-CavityOptomechanics,Eichenfield:2009-OptMechCrystals,WeisKippenberg:2010-OMIT,
Rocheleau:2009,Teufel:2011-Cooling,Chan:2011-LaserCooling},
trapping of phonons in
phononic bandgap cavities \cite{Eichenfield:2009-OptMechCrystals,Chan:2011-LaserCooling},
photon translation via phonons \cite{TianPainterConversion:2012},
and indirect qubit-phonon coupling
\cite{OConnell:2010-GroundState,KolkowitzHarrisLukin:2012-NVc}.
What is missing to complete the analogy for phonons is a non-linear element similar to an atom in cQED.

Such an element  is possible,
where an impurity transition  
in a crystal
(e.g., two-levels of a Si donor)
couples  {\em directly} to confined
phonons to form a hybridized state,  
which has been referred to as a phoniton (in analogy with a polariton) \cite{OurPhoniton-PRL107}.
The impurity-phonon interaction
can be large due to a large deformation potential:
$\langle \psi_{s'}| \hat{D}_{ij} | \psi_s\rangle \sim \mbox{eV}$ \cite{BirPikusBook}.
The previously proposed system utilizing an
Umklapp {\em valley} transition \cite{Smelyanskiy:2005,OurPhoniton-PRL107}
of a donor in Si, however, requires very high frequencies
(a few hundred GHz) and can be difficult to integrate with
other phonon components.
While other  impurities
such as
in diamond \cite{KolkowitzHarrisLukin:2012-NVc} or  in III-V semiconductors
can offer smaller frequencies,
a practical system in silicon would
be highly desirable  given recent
demonstrations of high-Q cavities in silicon nanostructures \cite{Chan:2011-LaserCooling,NewPainterHighQ:2012},
silicon's  investment
in materials quality, 
and compatibility with
CMOS technology and  silicon photonics.

\begin{figure}[t]  
\begin{centering}
\includegraphics[width=0.9\linewidth]{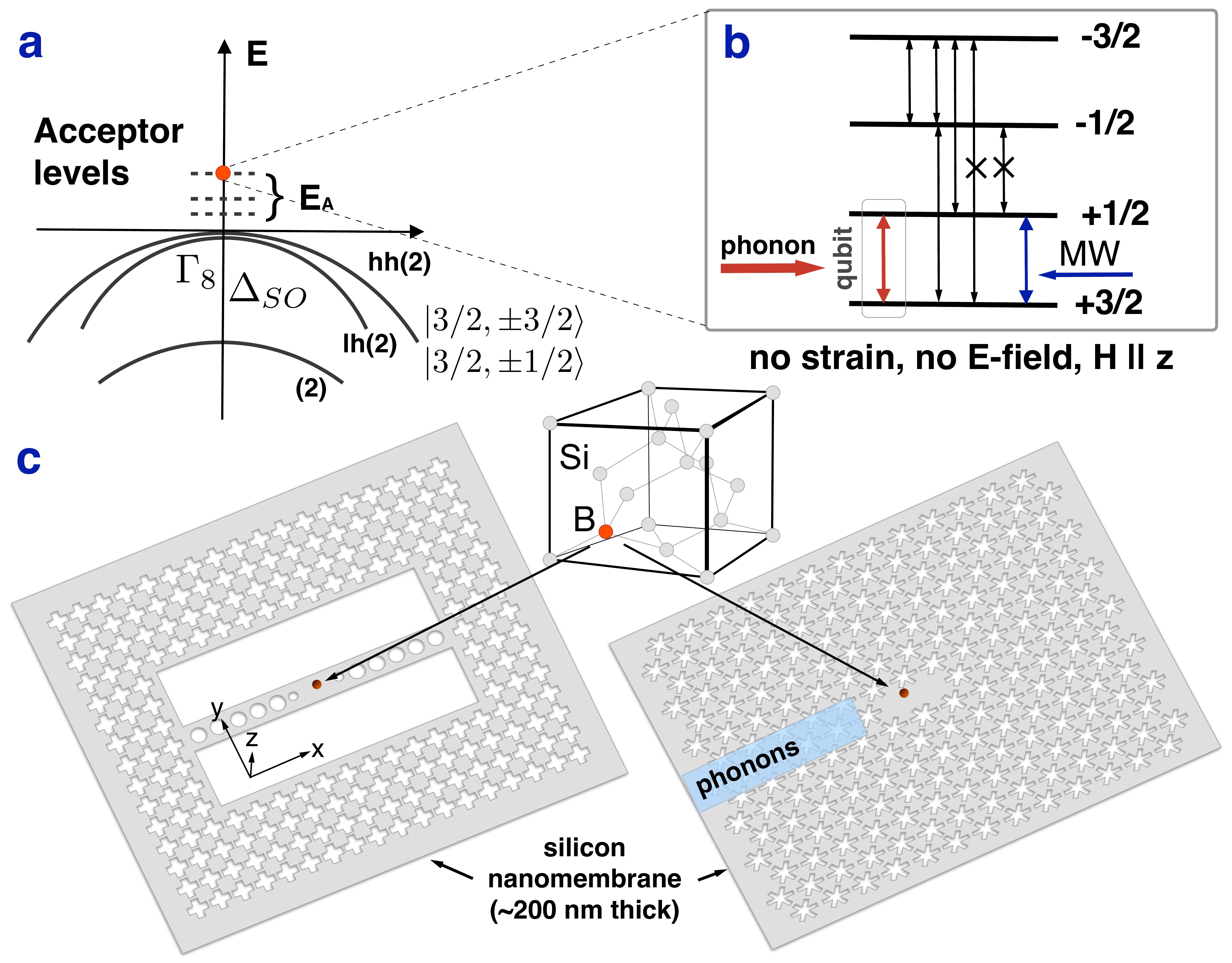}

\par\end{centering}
\caption{Acceptor:Si nanomechanical-cavity-phoniton.
(a) Hole valence bands in Si;
4-fold degeneracy  at the band top (and of lowest acceptor states)
corresponds to particles of spin $J=3/2$
($\Gamma_8$ representation of cubic symmetry, see, e.g. Ref.\cite{BirPikusBook}).
(b) Ground state splitting via external magnetic field along $[0,0,1]$ direction;
allowed (forbidden) phonon transitions and qubit phonon driving 
(see text).
Level rearrangement is via additional strain.
System manipulation via electric static/microwave fields  is possible.
(c)
Nanomechanical 1D and 2D phonon bandgap cavities reminiscent of already fabricated
high-Q cavities in a patterned Si membrane \cite{Eichenfield:2009-OptMechCrystals,Chan:2011-LaserCooling};
an on-chip phonon waveguide allows   
coupling to the  
phoniton system.
\label{fig:device}
}
\end{figure}

Here we propose a new quantum circuit element based on a single acceptor
(such as B, Al, In)
embedded in a patterned silicon nano-membrane,
and driven by a long wavelength phonon, $\lambda_{\rm phon} \gg a^*_{\rm acceptor} \sim {\rm few\, nm}$,
compatible with
opto{\bf /}mechanical components \cite{Eichenfield:2009-OptMechCrystals,Chan:2011-LaserCooling}.
The acceptor two-level system (qubit), Fig.\,1,
has already been proposed
for quantum computing \cite{GoldingDykman:2003} and
is easily tunable in the $1-50\, \mbox{GHz}$ range
by external magnetic field
and also by    
electric field or strain (allowing multiple qubit choices).
We show how the acceptor-cavity system
allows for both
strong resonant coupling (where the qubit-phonon coupling, $g$, is greater than
the loss mechanisms of the qubit and cavity, $\Gamma_{\rm qb}, \kappa_{\rm cav}$, respectively)
and strong dispersive coupling,
enabling the observability of
a phonon vacuum Rabi splitting,
and QND measurement of the cavity phonon number.
Experimental signatures of the system are given, via magnetic field and temperature dependence
(for $T \lesssim 1 \mbox{K}$),
utilizing optical techniques.

{\it Engineering the qubit levels.}
Two acceptor qubit arrangements are possible based on the lifting of the
4-fold   ground state degeneracy via external fields \cite{Supplementary}.  
For a magnetic field $\bm{H}_z=(0,0,H_z)$ along the  crystal $[0,0,1]$ growth direction
one can choose the lowest two Zeeman levels,
$|\phi_1\rangle = |3/2\rangle$, $|\phi_2\rangle = |1/2\rangle$,
as the qubit, which is
the primary  focus of this paper (Fig.\,1b).
The Zeeman type interaction is given by \cite{Luttinger:1956,BirPikusBook}:
$H_{\bm{H}} = \mu_B \{ g'_1 \bm{J} \bm{H}
+ g'_2 \left( J_x^3 H_x + \mbox{c.p.}  \right) \}$;
here c.p. is cyclic permutation of $x,y,z$; $J_{x},\ldots$, etc.  are  the
spin $3/2$ matrices
(in the crystal directions),
and
the renormalized  $g$-values $g'_1$, $g'_2$ ($\mu_B$ is Bohr magneton),
depending on the acceptor bound states,
fulfill  the relations
$|g_1'| \approx 1$, $|g_2'| \ll |g_1'|$ \cite{BirPikusBook,Neubrand1:1978,KopfLassmann:1992}.
The qubit   
splitting $\delta E_{H} \simeq \mu_0 g'_1 H$
is
tunable in the range
$\approx 1 - 40\, \mbox{GHz}$ for $H = 0.1 - 3\, \mbox{T}$.
The term $\sim g'_2 J_x^3$ lifts the
equidistancy:
the outer splittings (Fig.\,1b) are larger
than the middle one by $\frac{3g'_2}{g'_1} \simeq 0.09$.
For a field $\bm{H}$
tilted away from the crystal axis   
the qubit splitting is weakly angle dependent.

Alternatively, a second qubit arrangement involves  
mechanical stress in addition to the magnetic field.
Stress
lifts the ground state degeneracy only partially:
e.g., for stress along the
crystal  $\hat{z}$-direction (Fig.\,1c),
states $|\pm 3/2\rangle$ and $|\pm 1/2\rangle$ remain degenerate.
Providing the  stress causes
a splitting larger than
the  magnetic field splitting,
the levels   
in  Fig.\,1b
rearrange so that the lowest (qubit) levels  will be
$|\phi_1'\rangle = |-\! 1/2\rangle$, $|\phi_2'\rangle = |1/2\rangle$.
This forms an alternate {\em ``phonon protected''} qubit,
decoupled from phonons to first order
(the coupling can be switched on
via electric field, see below).
The effect of strain, $\epsilon_{\alpha\beta}$, is described by the Bir-Pikus  Hamiltonian\cite{BirPikusBook}:
\begin{eqnarray}
&& H_{\epsilon} = a'\, \mbox{Tr}\,\epsilon_{\alpha\beta} +
b'\, \epsilon_{xx}\,J_x^2  +  \frac{d'}{\sqrt{3} }\, \epsilon_{xy}\,\{J_x J_y\}_{+} + \mbox{c.p.}
\label{strain-Hamiltonian}  .
\end{eqnarray}
Experimentally \cite{Neubrand1:1978} the   
deformation potentials    
for B:Si    
are:
$b'\simeq -1.42\, \mbox{eV}$,  $d' \simeq -3.7\, \mbox{eV}$.
We estimate a splitting of $\delta E_{\varepsilon} \approx 1 - 10\, \mbox{GHz}$
for external stress   
of $10^5 - 10^6\, \mbox{Pa}$ \cite{stress}.
A larger stress  would
suppress the qubit-phonon coupling.

{\it Strong coupling of acceptor to confined acoustic phonon.}
Here, we  focus on the qubit
$\left\{1=|3/2 \rangle, 2=|1/2 \rangle\right\}$ that does not require static strain.
The coupling to a phonon mode is calculated
by adding a quantized phonon field \cite{Supplementary}  
 in addition  to  any classical field.
We consider  coupling
 to a plane wave   
$\varepsilon_{\rm vac}\,\bm{\xi}^{(\sigma)}_{\bm{q}}\, e^{-i\bm{q}\cdot\bm{r}}$
with polarization  $\bm{\xi}_{\bm{q}}^{(\sigma)}$
(transverse, $t_1,t_1$, or longitudinal, $l$)
and energy
$\hbar v_{\sigma} q$,
that proved to be a good estimation of coupling to modes with
realistic boundaries \cite{OurPhoniton-PRL107}.  
Moreover, the acceptor transition, unlike the \{P/Li\}:Si valley transition \cite{Smelyanskiy:2005,OurPhoniton-PRL107}, 
is less sensitive to the details of the confined phonon mode since the dipole approximation applies.
The   
matrix element
is proportional to  the ``phonon vacuum field'',  
$\varepsilon_{\rm vac}\equiv\left(\frac{\hbar\, q}{2\rho{\cal V}\, v_{\sigma} }\right)^{\!1/2}$;
$\rho$, ${\cal V}$, and $v_{\sigma}$ are the mass density, mode volume, and
 sound velocity    
in Si.
The coupling is 
(acceptor is placed at maximum {\em strain}  
unlike in \cite{OurPhoniton-PRL107} where the valley transition requires
placing at
maximum {\em displacement}):  
\begin{equation}
g^{3/2,1/2}_{\sigma}\! = d' \left(\frac{\hbar \omega_{12} }{8\rho \hbar^2 {\cal V} v^2_{\sigma} }  \right)^{\!1/2}
\left\{\!
\begin{array}{c}
 \cos\theta ,  \sigma=t_1\\
 i \cos2\theta  ,  \sigma=t_2\\
 -i \sin 2\theta ,  \sigma=l\\
\end{array}
\right\} e^{-i\varphi}
\label{coupling-phonon-12} ,
\end{equation}
where the polar angles, $\theta,\varphi$ of the wave vector $\bm{q}$ are  with respect to $\bm{H}||\hat{z}$.
Thus,
the mode $t_2$ has a maximum along the phonon cavity ($\theta\approx \pi/2$), Fig.\,1c.
An alternative is to have an in-plane magnetic field $\bm{H}_x$ along
the crystal $[1,0,0]$ $\hat{x}$-direction
(the latter is chosen to be along the   
cavity):
both modes $t_1,t_2$ (now at $\theta\approx 0$) are preferably coupled to the cavity.
The maximal coupling $g^{3/2,1/2}_{{\rm max},\sigma}$  scales as
$\propto \sqrt{q/{\cal V}}$,
as expected for a ($1s \to 1s$) transition.
For a cavity volume ${\cal V}\simeq d \lambda^2$
($d=200\, \mbox{nm}$ is the Si membrane  thickness)
we get
coupling in the range $g/2\pi \simeq 0.4 - 21\, \mbox{MHz}$ for
$1 - 14\, \mbox{GHz}$ (Table 1).
The other allowed
transition $|3/2 \rangle \to |\!-1/2 \rangle$
(at twice  the qubit frequency)
is  well detuned;   
the transitions $|3/2 \rangle \to |\!-3/2 \rangle$, $|1/2 \rangle \to |\!-1/2 \rangle$
are phonon forbidden (Fig.\,1b).

Generally, when the in-plane magnetic field has some angle
$\theta_0$  with the cavity (crystal $x$-axis),
all transitions are allowed.
Also, the qubit coupling
to a preferably confined cavity phonon will change.
As a qualitative example, consider a plane wave
transverse mode $t_1$ (or $t_2$)  along the $x$-axis ($\theta\approx 0$): the  coupling
will change in the same way as in Eq.~(\ref{coupling-phonon-12}),
with $\theta$ replaced by $\theta_0$.
This allows manipulation of  the qubit-cavity coupling by
rotation of the magnetic field.

\begin{table*}
{\footnotesize
\begin{tabular}{|l|l|c|c|c|c|c|c|c|}
\hline
parameter  & symbol  &  circuit-QED  & Quant Dot-QED {}& {\scriptsize B:Si} (1 GHz) & {\scriptsize B:Si} (4 GHz) & {\scriptsize B:Si} (8 GHz) & {\scriptsize B:Si} (1 Tesla) \tabularnewline
\hline
resonance freq. &\, $\omega_{\rm r}/2\pi$ & $5.7\,{\rm GHz}$ & $325\,{\rm THz}$   {}& $1\,{\rm GHz}$       & $4\,{\rm GHz}$  & $8\,{\rm GHz}$  & $14\,{\rm GHz}$ \tabularnewline
\hline
vac. Rabi freq. &\, $g/2\pi$     & $105\,{\rm MHz}$          & $13.4\,{\rm GHz}$  {}& $0.41\,{\rm MHz}$   & $3.27\,{\rm MHz}$ & $9.26\,{\rm MHz}$ & $21.4\,{\rm MHz}$  \tabularnewline
\hline
cavity lifetime &\, $1/\kappa,Q$ & $0.64\,{\rm \mu s}$, $10^{4}$ & $5.5\,{\rm ps}$, $1.2\,10^{4}$  {}& $15.9\,{\rm \mu s}$, $10^{5}$ & $4\,{\rm \mu s}$ & $2\,{\rm \mu s}$  & $1.14\,{\rm \mu s}$  \tabularnewline
\hline
qubit lifetime  &\, $1/\Gamma$     & $84\,{\rm ns}$            & $27\,{\rm ps}$     {}& $386.5\,{\rm \mu s}$ & $6\,{\rm \mu s}$  & $0.75\,{\rm \mu s}$  & $0.14\,{\rm \mu s}$  \tabularnewline
\hline
critical atom {\scriptsize\#}  &\, $2\Gamma\kappa/g^{2}$ & $\lesssim 8.6\,10^{-5}$ & $\lesssim 1.87$         {}& $\lesssim 4.9\,10^{-5}$ & $\lesssim 2\,10^{-4}$ & $\lesssim 3.9\,10^{-4}$ & $\lesssim 6.9\,10^{-4}$   \tabularnewline
\hline
crit.\! phonon {\scriptsize\#}  &\, $\Gamma^{2}/2g^{2}$  & $\lesssim 1.6\,10^{-4}$ & $\lesssim 9.4\,10^{-2}$ {}& $\lesssim 5.1\,10^{-7}$ & $\lesssim 3.2\,10^{-5}$ & $\lesssim 2.6\,10^{-4}$ & $\lesssim 1.4\,10^{-3}$  \tabularnewline
\hline
{\scriptsize\#} Rabi flops  & ${\scriptsize 2g/(\!\kappa\!+\!\Gamma)}$  &  $\sim 98 $  & $\sim 0.8$  {}& $\sim 79$ & $\sim 99$ & $\sim 64$ & $\sim 34$  \tabularnewline
\hline
{} cavity volume & ${\cal V}$  & $10^{-6}\,\lambda^3$  & -  {}& $0.037\,\lambda^3$ & $0.148\,\lambda^3$ & $0.296\,\lambda^3$ & $0.52\,\lambda^3$  \tabularnewline
\hline
{} wavelength  & $\lambda$  &  5.26\,{\rm cm}  & $921\,{\rm nm}$  {}& $5400\,{\rm nm}$ & $1350\,{\rm nm}$ & $675\,{\rm nm}$ & $385\,{\rm nm}$  \tabularnewline
\hline
{} dispersive coupling  & $\chi\equiv{\scriptsize g^2/\Delta}$  &  $17\,{\rm MHz}$  & --  {}& $0.04\,{\rm MHz}$ & $0.33\,{\rm MHz}$ & $0.93\,{\rm MHz}$ & $2.14\,{\rm MHz}$  \tabularnewline
\hline
{} peaks' resolution      & ${\scriptsize 2\chi/\Gamma}$  & $\sim 6 $ & -- {}& $\sim 199$ & $\sim 25$ & $\sim 9$  & $\sim 4$  \tabularnewline
{\scriptsize\#} of peaks & ${\scriptsize 2\chi/\kappa}$  & $\sim 70$ & -- {}& $\sim 8$   & $\sim 16$ & $\sim 23$ & $\sim 31$  \tabularnewline

\hline
\end{tabular}
}
\caption{
Key   
parameters for
circuit-QED \cite{Schuster:2007-PhotonNumberStates}
(1D cavity), Quantum dot(QD)-QED \cite{E_Waks:2012}  
vs. the 
$\{|3/2 \rangle, |1/2 \rangle\}$ B:Si phoniton   
in a patterned Si membrane (of thickness $d=200\,\mbox{nm}$) phononic bandgap cavity;
we show calculations for maximal coupling at frequencies of $1\,\mbox{GHz}$,
$4\,\mbox{GHz}$, $8\,\mbox{GHz}$, and $14\,\mbox{GHz}$,
for cavity volume ${\cal V}=d\lambda^2$  and  $Q=10^{5}$,
using bulk $T_1$-limited linewidth $\Gamma$.
The limiting frequency for strong dispersive coupling is reached at $\approx 21\,\mbox{GHz}$,
when $\chi = \Gamma$; dressed state's resolution parameters are comparable to that in circuit-QED.
\label{tab:Key-rates}
}

\label{table:parameters}
\end{table*}

{\it Qubit relaxation rate.}
The qubit relaxation in the cavity is bounded at low temperatures by the bulk phonon spontaneous emission rate, we find:
\begin{eqnarray}
&&\Gamma_{3/2,1/2}(\theta_0) = \frac{(\hbar \omega_{12})^3}{20\pi\rho \hbar^4}\, \left\{ d'^2 (\cos^2{2\theta_0} +1)
\left[ 2/3 v^5_l + 1/v^5_t  \right] \right.
\nonumber\\
&& \qquad\qquad\quad \left. {} + b'^2\,\sin^2{2\theta_0} \left[ 2/v^5_l + 3/v^5_t \right]  \right\}
\label{emission-rate-12} ;
\end{eqnarray}
here $l$-phonon contribution is a small percent.   
The results in Table~1 are for $\theta_0 = 0$.
   Note that the coupling in this case can be switched off
   (e.g. for a $t_1$-mode along $\hat{x}$-direction, at $\theta_0=\pi/2$) while the relaxation cannot.

For the alternate qubit, $\{|\!-\! 1/2\rangle, |1/2\rangle\}$,
the stress and magnetic field
are parallel along the
$\hat{z}$-direction (Fig.\,1c).
Here both coupling and relaxation are zero in the absence of electric field and can be switched on
using non-zero  electric field $\bm{E}_z$ in the same direction \cite{Supplementary}.  
The qubit-phonon coupling is given by the same
Eq.(\ref{coupling-phonon-12}) multiplied by a coupling factor,
a function of the splitting ratios $r_h \equiv \frac{\delta E_{H}}{\delta E_{\varepsilon}}$,
$r_e \equiv \frac{\delta E_{E}}{\delta E_{\varepsilon}}$:
$f(r_h,r_e)=(\sqrt{z_{+}z_{-} } -1)/\sqrt{(1+z_{+})(1+z_{-})}$,
with $z_{\pm}=(1\pm \sqrt{(1 \mp r_h)^2 + r_e^2} \mp r_h)^2/r_e^2$.
Thus, e.g. for $r_h=0.5-0.9$
this factor  reaches $\approx 0.25-0.65$ for some optimal value
of the electric field splitting, $r_e \lesssim 1$, which allows strong  
coupling.

The calculated relaxation times from Eq.(\ref{emission-rate-12})  
(Table~1) are comparable to  that 
in bulk Si \cite{Golding:2011} at low  B:Si doping ($8\times 10^{12}\, \mbox{cm}^{-3}$
or $500\, \mbox{nm}$  acceptor spacing),
where    
$T^{\rm echo}_1 \simeq 7.4\,\mu\mbox{s}$  and $T^{\rm echo}_2 \simeq 2.6\,\mu\mbox{s}$
were measured  at $45\, {\rm mK}$.
Note that the single acceptor linewidth ($\sim 1/T_{2,\rm single}^*$)
is the proper metric to compare $g$ with,
not the inhomogeneously broadened $1/T_2^*$ obtained from ensemble  measurements
\cite{Tezuka:2010-IsotopPurifiedEPR-BSi}.
While  $T_{2}=2 T_1$  for     
phonons alone, it may be limited by
electric-dipole
coupling to impurities \cite{GoldingDykman:2003},
magnetic hyperfine coupling to nearby nuclei ${}^{29}Si$ (expected to be small for holes),
or charge noise (though
here the acceptor is far away from surfaces or metal gates);
both $T_1$, $T_2$ may  
improve for defect-free, low-doped,
\cite{Neubrand1:1978,KopfLassmann:1992,StegnerAPL:2011}
and isotopically purified  samples
\cite{Karaiskaj:2003-IsotopAcceptorSplit,Tezuka:2010-IsotopPurifiedEPR-BSi};
$T_1$ may also improve in nanomembranes ($d\ll\lambda$) due to phase-space suppression
(less available modes).

{\it Phonon cavity loss.}
In the 1D/2D-phononic bandgap Si nanomembranes considered here
(Fig.\,1c),
the main cavity loss mechanism is due to (fabrication) symmetry-breaking effects,
coupling the cavity mode to unconfined modes,
and also due to cavity surface defects \cite{NewPainterHighQ:2012}.
Bulk losses
are negligible in the few GHz
range \cite{OurPhoniton-PRL107,NewPainterHighQ:2012}.
In this range the cavity $Q$-factor, $Q\equiv \kappa/\omega$,
can reach
$10^4 - 10^5$, or
higher \cite{Chan:2011-LaserCooling,NewPainterHighQ:2012}.

Calculated rates  in Table\,1
show that {\it strong resonant coupling}
is possible:
$g_{12}  \gg \Gamma_{12}, \kappa$
in a wide frequency range,   
allowing $\sim 30-100$  Rabi flops.
The low   
limit of
$1\,\mbox{GHz}$ is for $T\simeq 20\,\mbox{mK}$,
unless an active cavity cooling is performed \cite{Chan:2011-LaserCooling};
a high limit of $\sim 200\, \mbox{GHz}$ is set by the
different energy scaling of $g$ and $\Gamma$.
At high frequencies the $Q$-factor will decrease;
still, e.g., at
$14\, \mbox{GHz}$
even $Q=10^3$ leads to strong coupling.

\begin{figure*}[t]  
\includegraphics[width=0.76\linewidth]{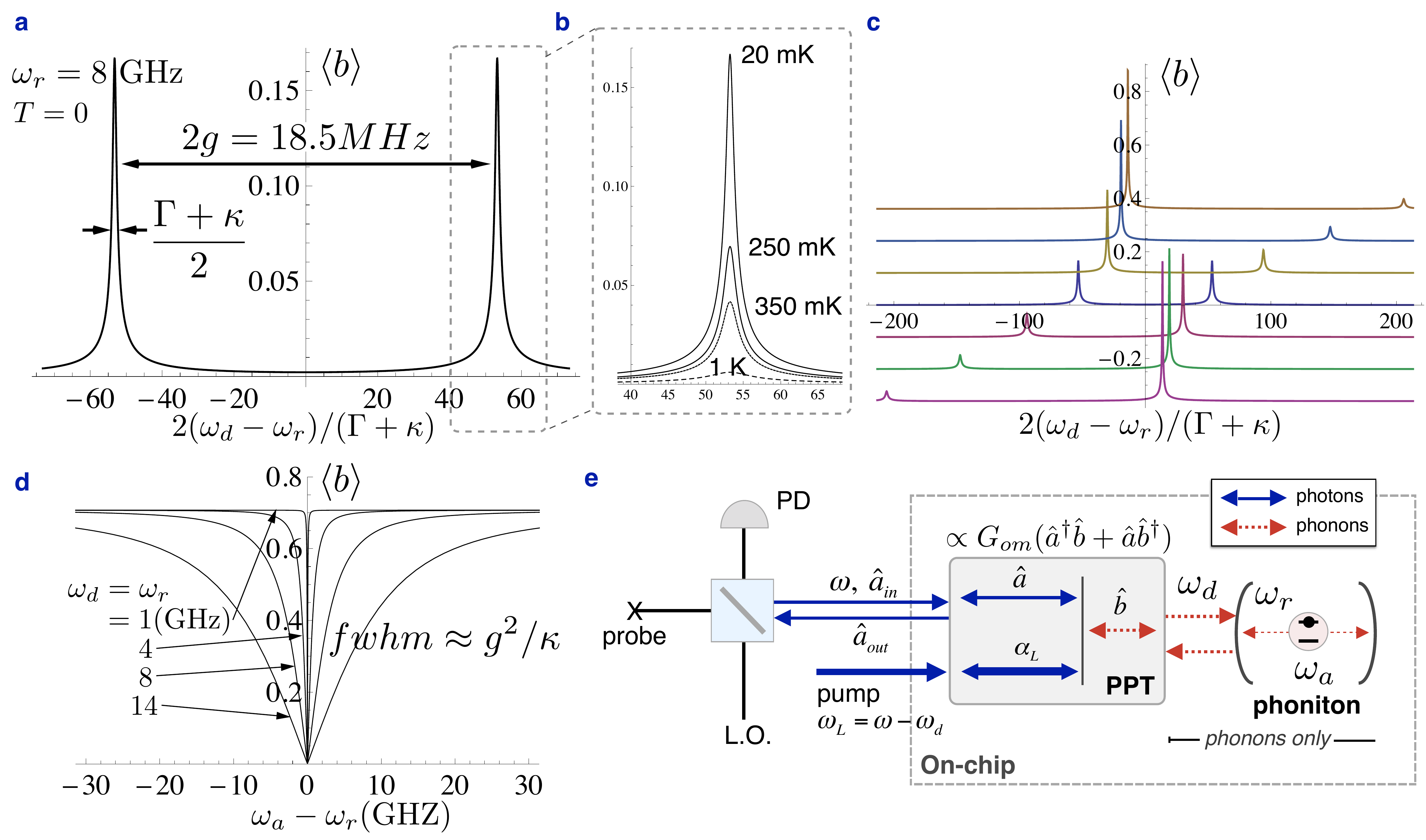}

\caption{
Intracavity field $\langle \hat{b}\rangle$ vs. frequencies and temperature.
(a) Rabi splitting in the strong resonant coupling regime as a function of a
phonon   
signal sweep.
(b) Rabi peaks  vs.\,  temperature (approximate).
At $n_{\rm th} \gtrsim 1$ broad peaks will appear inside the Rabi doublet (not shown)
due to transitions to higher dressed states (see text).
(c) ``Anticrossing picture'' at qubit-phonon cavity detuning, $\Delta_{\rm ar}$ (curves vertically shifted for clarity);
for large positive detuning (upmost curve),  the  left resonance is dispersively shifted by $-g^2/\Delta_{\rm ar}$
while the  
right resonance is   
suppressed as $\sin \Theta_0$ (see text).
(d) An alternative   
to determine the Rabi splitting: the absorption spectrum as
a function of qubit detuning (via a magnetic field sweep).
(e) Schematics of 
an optical  homodyne/heterodyne
experiment
(see also Ref.\,\cite{WeisKippenberg:2010-OMIT})
utilizing a PPT  \cite{SafaviNaeini:2011-Translator,Supplementary}.
\label{fig:experiment}
}
\end{figure*}

{\it Observing the vacuum Rabi splitting.}
A suitable observable   
is the averaged phonon cavity field amplitude $|\langle \hat{b} \rangle|$,
which we have calculated, Fig.\,2a-d,
taking into account the first two excited dressed states
(a ``two-state approximation'', \cite{TianCarmichael:1992}).
On Fig.\,2a we show the 
$\langle \hat{b} \rangle$-spectrum ($\omega_r$ is the cavity frequency)
for $\omega_a = \omega_r = 8\, \mbox{GHz}$
as two Rabi peaks at $\pm g$
vs.\ external phonon driving frequency $\omega_d$,
while Fig.\,2c shows the ``anticrossing picture'' of the spectra at different
acceptor  detuning
$\Delta_{\rm ar} = \omega_a - \omega_r$.
Here the left resonance width and height 
are
$\Gamma_L =\sin^2 \Theta_0 \Gamma + \cos^2 \Theta_0 \kappa$ and $\cos \Theta_0$,
and for the right resonance one replaces  
$\sin \Theta_0 \leftrightarrow \cos \Theta_0$,
with $\tan (2\Theta_0) = 2g/\Delta_{\rm ar}$.

Another   
option to observe the  dressed resonance(s) is to sweep the qubit detuning
(via the $\bm{H}$-field)  while keeping the 
input in resonance with the phonon cavity.  
As seen from
Fig.\,2c,
the field amplitude will exhibit a resonant dip, 
shown
for  
different cavity frequencies, 
Fig.\,2d.
For large detuning the resonance is approximately a Lorentzian   
with a full width at half maximum, $\{\mbox{fwhm}\} \approx g^2/\kappa$
(a weak dependence on the qubit relaxation $\Gamma$  is suppressed for strong coupling).

With increasing temperature the Rabi peak
will be broadened \cite{MilburnWalls}
by the factor $1+2 n_{\rm th}$
and the peak height will decrease by
$2 p_{\rm st} -1$,
where $p_{\rm st}=(1+n_{\rm th})/(1+2 n_{\rm th})$ is the ground state occupation.  
This is relevant for small thermal phonon number,
$n_{\rm th}\equiv 1/(\exp\left[\hbar\omega_r/k_B T\right]-1)$,
at low $T$
($n_{\rm th}\lesssim 0.2$; see, e.g., Ref.\,\cite{WallraffTemp}).
For $\omega_r/2\pi = 8\,\mbox{GHz}$ the Rabi peaks will be seen at $350\, \mbox{mK}$,
but are negligible at 1K (Fig.\,2b).
At higher $T$,   
when $n_{\rm th}\gtrsim 1$, the lowest dressed states
become saturated and in addition
two broadened peaks will appear (inside the  Rabi doublet, Fig.\,2a),
at
$\delta\omega_d \simeq \pm g[\sqrt{n+1}-\sqrt{n}],\, n\approx n_{\rm th}$.
These   
will dominate    over the Rabi peaks \cite{WallraffTemp},
providing     
a signature of  strong coupling even beyond 1K.

{\it Measurement via photons and standard techniques.}
Ideally, one would probe the acceptor-cavity system with phonons.
Direct phonon creation and detection   
should be possible via  acoustic transducers \cite{Delsing:2012-SAW}.
As in circuit QED \cite{Wallraff:2010-antibunching}, phonon correlations
can be measured
even without   
single phonon counters.
However, here we consider an approach
with single phonon sensitivity
using
a phonon-to-photon translator (PPT) \cite{SafaviNaeini:2011-Translator}
that can be realized on
the same nanomembrane   
(with a   
photon/phonon bandgap, see Figs.\,1c,\,2e).\

The PPT allows for
optical techniques  \cite{WeisKippenberg:2010-OMIT,Chan:2011-LaserCooling}
to be applied to phononics. 
We show in Fig.\,2e an experimental schematic,
to measure
the phonon cavity field  
$\langle \hat{b} \rangle$
via a homodyne/heterodyne optical measurement \cite{MilburnWalls}.
To  
scan around the mechanical resonance $\omega_r$,
one needs optical frequency resolution
(at $\omega/2\pi \approx 200\,\mbox{THz}$)
better
than the dressed state  
width,
$(\Gamma+\kappa)/2\pi \approx  30-150\,\mbox{kHz}$, Table\,1.
In the on-chip  PPT implementation
no photon    
should enter the phoniton
system,
to avoid acceptor ionization.
However,
the estimated ionization cross section is small, $\sigma_{\rm phot} \approx 8.6\times 10^{-23}\, {\rm m}^2$:
for 10 photons in the cavity, at maximum photon-acceptor overlap,
one gets an   
ionization lifetime of $12\,\mu{\rm s}$,
that can be further increased
in regions of low photon intensity  \cite{Supplementary}.

A strong {\it dispersive coupling}, $\chi \equiv g^2/\Delta_{\rm ar}$, is
reachable  as per Table\,1.
Since  $\Delta_{ar} \geq 10\, g$  (dispersive regime)
and $g^2/\Delta_{ar}\Gamma_{\rm relax} \geq 1$ (good resolution of phonon numbers),
resolving the
number states $|n\rangle$ in the phononic cavity would be possible.
In the dispersive regime one can apply
two tones (as in circuit-QED \cite{Schuster:2007-PhotonNumberStates}):
here, one tone is a {\it phonon}  probe  at $\omega_d$, slightly detuned
from the resonator  (Fig.\,2e).
The second (spectroscopic) tone,
at $\omega_{sp}$,  is driving some of the dressed transitions
around the acceptor frequency, at $\omega_a+(2\,n+1)\chi$, via electric microwaves (MW),
similar to
a manipulation of individual nitrogen-vacancy centers\cite{KolkowitzHarrisLukin:2012-NVc}
(for B:Si a stronger MW coupling \cite{KopfLassmann:1992,Golding:2011} is expected compared to NVc).
Thus, one could observe the {\em fine spectral structure}
of the dressed cavity-acceptor system,
predicted by Ref.\cite{DykmanKrivoglaz} in a different context,
by measuring the phonon (photon) reflection
while sweeping
the MW tone $\omega_{sp}$.
We note that  other measurement approaches are possible
via, e.g.,
hole transport \cite{Rogge:2004-StarkSi,Calvet:2007-LinearStark} or
STM probe spectroscopy \cite{SLoth:2010-STM-SpinControlElCurrent}.

{\it Discussion and applications.}
We have introduced a
system that allows for on-chip manipulation of coherent acoustic phonons via
coupling to acceptor qubit  states
in a nanomechanical cavity.
Hybridization of the phonon-acceptor system
and strong dispersive coupling are
possible with comparable parameters to   
circuit-QED \cite{Schuster:2007-PhotonNumberStates}
and far surpassing semiconductor
QD QED \cite{E_Waks:2012}.
The cavity-phoniton
can be incorporated in more complex  networks
such as with phonon-photon
interfaces to photonics \cite{Chan:2011-LaserCooling,SafaviNaeini:2011-Translator},
and in arrays of phonitons
for engineered many-body phonon devices \cite{Rabl:2010-indirect,OurPhoniton-PRL107}.
From the perspective of qubits
\cite{GoldingDykman:2003,Morello:2012-SingleAtomSi},
the isolated acceptor provides a potentially robust
two-level system
for quantum information  processing.
Our system offers an avenue for phonon
dispersive readout of acceptor qubits
and the potential for spin qubit-to-photon conversion
in silicon.

\section*{Supplemental Material}

{\it Hole valence bands in Si.}
Holes in Si require a richer physical picture \cite{Luttinger:1956,BirPikusBook}
(compared to positrons in QED).
The 4-fold degeneracy (Fig.\,1a, main text) at the top of the valence band  (neglecting heavy-light hole splitting)
corresponds to propagation of particles of spin $J=3/2$,
reflecting the $\Gamma_8$ representation of cubic symmetry.
Relatively large spin-orbit coupling implies a 2-fold degenerate band
($\Gamma_7$ representation), split-off by an energy gap
$\Delta_{SO} \simeq  45\, \mbox{meV}$ \cite{BirPikusBook}.
The role of an ``atom'' can be taken by a single impurity in a
host crystal, e.g. in Si.    
For shallow acceptor centers in Si (e.g., B, Al, In, etc)
the ionization energy is  $E_A \sim \Delta_{SO}$: thus, all valence bands
will play a role in the acceptor states.
Still, the lowest acceptor states
remain 4-fold degenerate
since the acceptor spherical (Coulomb) potential does not change the
cubic symmetry of the host crystal \cite{BirPikusBook}.

{\it Origin of strong acceptor-phonon coupling.}
The impurity-acoustic phonon interaction \cite{BirPikusBook},
$H_{\rm e,ph}^{\rm ac}(\bm{r})=\sum_{ij}\hat{D}_{ij}\,\hat{\varepsilon}_{ij}(\bm{r})$, may lead to
a strong coupling regime  ($g > \Gamma_{\rm qb},\kappa_{\rm cav}$)
even for  cavity effective volume of few tens \cite{OurPhoniton-PRL107} of $\sim \lambda^3$,
since the deformation potential
matrix elements are large \cite{BirPikusBook}:
$\langle \psi_{s'}| \hat{D}_{ij} | \psi_s\rangle \sim \mbox{eV}$.
Qualitatively, the large coupling can be traced from the much smaller  bandgap ($\sim \mbox{eV}$)
in the ``Si-vacuum'' as compared to QED
($\sim 10^6\mbox{eV}$).
For the spin transitions of interest (e.g., $3/2 \to 1/2$) the
spin states are actually  ``compound'' states of  $P$-like Bloch orbitals (spin $1$) and electronic spin-$1/2$.
Thus, coupling via deformation (a phonon) of the crystal   is possible due to different
orbital content of these states.

{\it Engineering the qubit levels. The role of electric field.}
The degeneracy of the ground state can be lifted by
external magnetic field
via the Zeeman type interaction $H_{\bm{H}}$ (Fig.\,3a),
and via mechanical strain (Fig.\,3b) (see Eq.~(1) of main text).  
The related Hamiltonians are invariants
of the cubic symmetry group  $O_h = T_d \times I$
and time reversal \cite{Luttinger:1956,BirPikusBook}
and are constructed from the
momentum operator,
$k_\alpha = \frac{1}{i}\frac{\partial}{\partial x_\alpha} + \frac{e}{c} A_\alpha$
(or the corresponding fields \cite{BirPikusBook})
and the spin-$3/2$ operators $J_{\alpha},\, \alpha=x,y,z$.

\begin{figure}[t]  
\includegraphics[width=\linewidth]{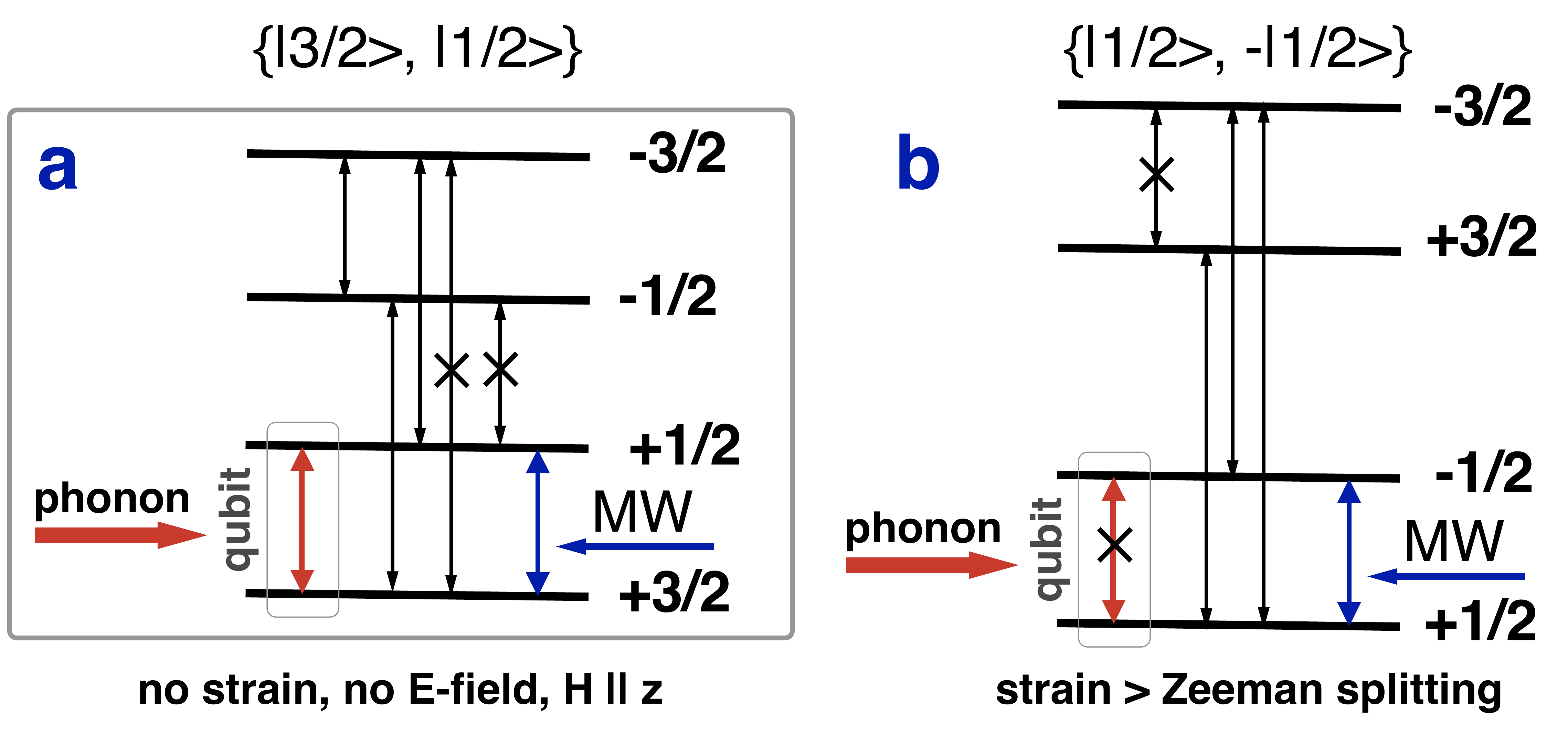}

\caption{
Two possible level arrangements as described in main text.
(a) Splitting due to magnetic filed along a crystal direction (e.g. $\vec{H}||\hat{z}$).
Equidistancy is lifted by the cubic term, leading the 9\% smaller  (for $g_1 g_2 < 0$)
splitting of the levels $-1/2$, $1/2$. The qubit, $\{|1/2\rangle, |3/2\rangle\}$,
is coupled strongly to a confined phonon, and can be manipulated via MWs.
(b) Level splitting in presence of magnetic field and stress along the $\hat{z}$-direction,
in the case of $\delta E_{\varepsilon} > \delta E_{\bm{H}}$.
The alternate qubit, $\{|\!-\! 1/2\rangle, |1/2\rangle\}$, is decoupled from phonons
(to first order). The coupling can be switched on via suitable electric field in the same direction.
\label{fig:levels1-2}
}
\end{figure}

For a relatively weak electric field $\bm{E}$ the linear Stark effect is possible:
\begin{equation}
H_{\bm{E}} = \frac{p_E}{\sqrt{3}} \left( E_x \{J_y J_z\}_{+} + E_y \{J_z J_x\}_{+} + E_z \{J_x J_y\}_{+}  \right)
\label{linear-Stark} ,
\tag{S1}
\end{equation}
since an ion impurity actually reduces the cubic symmetry ($T_d \times I$) to $T_d$
(and thus, there is no invariance under inversion) \cite{BirPikusBook}.
The ground state splits to two doubly degenerate levels; however, the $H_{\bm{E}}$ does not commute
with $J_z$ for any direction of the field $\bm{E}$, leading to mixing of the Zeeman states.
The latter can be useful to switch on/off the phonon coupling of the alternate qubit,
$\{|\!-\! 1/2\rangle, |1/2\rangle\}$ (Fig.\,1b),
provided the splitting $\delta E_{\bm{E}} = 2 p_E |\bm{E}|$ is of the order of that due to stress,
e.g., in the GHz range.
The transition electric dipole moment, $p_E$, can be extracted
from experiments: bulk dielectric absorption measurements \cite{KopfLassmann:1992}
give  $p_E \simeq 0.26\, \mbox{D}$,
with $D=3.336\times 10^{-30}\, \mbox{C m}$ being the Debye unit for e.d.m.
(this is supported by single acceptor transport experiments \cite{Calvet:2007-LinearStark}).
Thus, a splitting  of $1\, \mbox{GHz}$ requires
an  electric filed $|\bm{E}|_{\rm 1Ghz} \simeq 3.85\times 10^{5}\, \mbox{V/m}$,
achievable in nanodevices \cite{Rogge:2004-StarkSi}.
Note, however, that increasing the field (splitting) exponentially decreases the qubit life time
due to acceptor ionization: for $\delta E = 1\,\mbox{GHz}$ the life time is $\tau_{\rm ion}\approx 12\,{\rm s}$,
while for $\simeq 1.26\,\mbox{GHz}$ it is $\tau_{\rm ion} \approx 12\,{\rm ms}$, etc \cite{Rogge:2004-StarkSi}.

These numbers show that there is an experimental ``window'' for the alternate  qubit,
$\{|\!-\! 1/2\rangle, |1/2\rangle\}$, introduced in the main text.
For example, for a qubit (Zeeman) splitting of $\delta E_{\bm{H}} = 1\,\mbox{GHz}$
and a strain splitting $\delta E_{\varepsilon} = 1.43\,\mbox{GHz}$ (ratio of $r_h \equiv\frac{\delta E_{\bm{H}} }{\delta E_{\varepsilon} }=0.7$)
the coupling factor  reaches the maximal value of $f(r_h,r_e)\simeq 0.4$ for $\delta E_{\bm{E}} = 1\,\mbox{GHz}$, i.e., $r_e = 0.7$.
Analogously, for a qubit splitting of $\delta E_{\bm{H}} = 2\,\mbox{GHz}$   this electric field splitting leads to the same coupling factor of $0.4$,
giving a possibility for a strong acceptor-phonon coupling (Eq.~(2) of the main text),  
and a relatively long
(static field) ionization life time.

{\it Strong coupling of acceptor to confined acoustic phonon.}
We account for the acceptor coupling to a quantized phonon field starting from the
Bir-Pikus
Hamiltonian, derived for a uniform classical strain field, Eq.~(1) (see main text). 
For low-energy acoustic phonons the
interaction Hamiltonian,  $\hat{H}_{\rm ph}$,
has the same  form   
with  the  strain operator
$\hat{\varepsilon}_{ij}(\bm{r})=\frac{1}{2}\left(\frac{\partial u_{i}}{\partial r_{j}}+\frac{\partial u_{j}}{\partial r_{i}}\right)$
expressed via the  quantized
mechanical displacement:
$\bm{u}(\bm{r})=\sum_{\bm{q},\sigma}\left(\bm{u}_{\bm{q}\sigma}(\bm{r})\, b_{\bm{q}\sigma}+
\bm{u}_{\bm{q}\sigma}^{*}(\bm{r})\, b_{\bm{q}\sigma}^{\dagger}\right)$.
The mode normalization
is
$\int d^{3}\bm{r}\,\bm{u}_{\bm{q}\sigma}^{*}(\bm{r})\bm{u}_{\bm{q}\sigma}(\bm{r})=\frac{\hbar}{2\rho\,\omega_{\bm{q}\sigma}}$,
so that  $b_{\bm{q}\sigma}^{\dagger}$   creates a phonon in the mode
$(\bm{q},\sigma)$  with energy  $\hbar\omega_{\bm{q}\sigma}$       
($\rho$ is the material mass density)
in a mode volume ${\cal V}$.
The vector $\bm{q}$  denotes a collective
index of the discrete phonon mode defined via the phonon cavity  boundary
conditions and
mode volume ${\cal V}$.
Similar to  cavity QED   \cite{MilburnWalls},
the phonon-acceptor coupling $\hbar g^{s's}_{\bm{q}\sigma}\equiv \langle s';\bm{q}\sigma | H_{\rm ph} | s\rangle$
enters in a Jaynes-Cummings
Hamiltonian (see, e.g. Ref. \cite{OurPhoniton-PRL107}):
$H_{\rm g}\approx \hbar g^{s's}_{\bm{q}\sigma}\left(\sigma_{s's}^{+}b_{\bm{q},\sigma}+\sigma_{s's}^{-}b_{\bm{q},\sigma}^{\dagger}\right)$,
where we only retain the resonant cavity phonon,
and $\sigma_{s's}^{+}\equiv|s'\rangle\langle s|$
refers to  the relevant acceptor  transition.

{\it Phonon-photon translator.}
Such a device is based on optomechanical  non-linearities that couple  
in the same bandgap cavity (see Fig.\,2e of main text)
two photon modes ($\hat{a}, \hat{a}_p$) and a phonon mode $\hat{b}$,
via  optomechanical coupling $h_{om}$ \cite{SafaviNaeini:2011-Translator}.
For photons in the near-infrared range ($\lambda_{opt} \approx 1500\, \mbox{nm}$)
the PPT allows one to couple  
a quantum optical input/output channel (of frequency $\omega/2\pi\simeq 200\, \mbox{THz}$)  to a phonon channel
(with $\omega_d/2\pi \simeq 4-8\, \mbox{GHz}$),
and the coupling between the fields is enhanced
by the auxiliary photon pump channel,
pumping at the   sideband resolved
frequency $\omega_p =\omega - \omega_d - \Delta$
(at pump detuning $\Delta=0$ it is at resonance with the red   
sideband of mode $\omega$).
The coherent nature of the photon-to-phonon translator is described by the effective
beam-splitter type Hamiltonian \cite{Braunstein:2003-Transfer}:
\begin{equation}
H_{\rm b-s} = -\Delta \hat{b}^{\dagger} \hat{b} + G_{\rm om} \left( \hat{a}^{\dagger} \hat{b} + \hat{a} \hat{b}^{\dagger} \right)
\label{beamsplitter}
\tag{S2}
\end{equation}
where $G_{\rm om} \propto h_{om} E_0$ is the enhanced effective coupling,
proportional to the pump field amplitude, $E_0$.
The  weak coupling regime, $G_{\rm om} < \kappa^{opt}$  is needed to avoid total optical reflection,
and  optimal  translation (close to $100\%$) takes place at a matching condition \cite{SafaviNaeini:2011-Translator}
$G_{\rm om} = \sqrt{\kappa^{opt} \kappa^{mech}}$
($\kappa^{opt}$, $\kappa^{mech}$, are the couplings of the PPT to respective photon/phonon waveguides).

{\it Acceptor ionization via optical photons.}
Since the photon-to-phonon translator is realized on the same Si nanomembrane
(implying a simultaneous  photonic/phononic bandgap structure) it is natural to
ask how  $200\, \mbox{THz}$ photons  may affect the qubit lifetime when they reach the acceptor.
The corresponding photon energy of $0.82\,\mbox{eV}$ is less than the indirect bandgap in Si
($\Delta E_{\rm gap} =1.1\,\mbox{eV}$) and thus interband transitions are not possible.
Thus, one considers an ``ionization process'' of a bound hole going to the continuous spectrum,
an analog of the ionization of an (anti)hydrogen atom
(the corresponding cross section is thus  re-scaled).
Correspondingly, one uses the re-scaled values: a free hole mass $m_A \simeq 0.23\,m_e$ in Si,
an effective Bohr radius $a^{\rm eff}_A = \frac{e^2 Z}{2 [4\pi\varepsilon_0 \varepsilon_r] E_A}$,
with the acceptor ionization energy for B:Si, $E_A \approx 0.044\,\mbox{eV}$,
and screening factor $Z\simeq 1.4$.
The total cross section is:
\begin{equation}
\sigma_{\rm phot} = \frac{32\pi}{3} \frac{\hbar^6}{c\sqrt{2 m_A} m_A^3 [a^{\rm eff}_A]^5 E_f^{2.5} (E_f+E_A)}
\frac{1}{[4\pi\varepsilon_0 \varepsilon_r]}
\label{ionization} ,
\tag{S3}
\end{equation}
where $E_f = \hbar\omega - E_A$ is the final (free) hole energy, and
$c = c_0/\sqrt{\varepsilon_r}$ is the speed of light in Si ($\varepsilon_r^{\rm Si} \simeq 11.9$).
Since $E_A \ll E_f$, the total cross section is suppressed as $\propto 1/E_f^{3.5}$ (final state energy suppression).
Given $n_c$ photons in cavity volume ${\cal V}\simeq d \lambda^2$  the acceptor life time is
$\tau_{\rm phot} = 2 {\cal V}/(n_c c \sigma_{\rm phot})$ for a maximum photon-acceptor overlap.
This limits the ability to perform active photon (sideband) cooling
of the phononic cavity (similar to Ref. \cite{Chan:2011-LaserCooling}).
However, by placing the acceptor close to a node of the photon cavity,
the ionization life time can increase considerably.


\begin{thebibliography}{10}

\bibitem{Kimble:1998}
H.~J. Kimble,
\newblock Physica Scripta {\bf T76}, 127 (1998).

\bibitem{Raimond:2001}
J.~M. Raimond, M.~Brune, and S.~Haroche,
\newblock Rev Mod Phys {\bf 73}, 565 (2001).

\bibitem{Blais:2004}
A.~Blais {\em et~al.},
\newblock Phys. Rev. A {\bf 69}, 62320 (2004).

\bibitem{Houck:2007-SinglePhoton}
A.~A. Houck {\em et~al.},
\newblock Nature {\bf 449}, 328 (2007).

\bibitem{Kippenberg:2008-CavityOptomechanics}
T.~J. Kippenberg and K.~J. Vahala,
\newblock Science {\bf 321}, 1172 (2008).


\bibitem{Eichenfield:2009-OptMechCrystals}
M.~Eichenfield  {\em et~al.},
\newblock Nature {\bf 462}, 78 (2009).


\bibitem{WeisKippenberg:2010-OMIT}
S.~Weis {\em et~al.},
\newblock Science {\bf 330}, 1520 (2010).

\bibitem{Rocheleau:2009}
T.~Rocheleau {\em et~al.},
\newblock Nature (London) {\bf 463}, 72 (2010).

\bibitem{Teufel:2011-Cooling}
J.~D. Teufel {\em et~al.},
\newblock Nature (London) {\bf 475}, 359 (2011).

\bibitem{Chan:2011-LaserCooling}
J.~Chan {\em et~al.},
\newblock Nature {\bf 478}, 89 (2011).



\bibitem{TianPainterConversion:2012}
C.~Dong~{\em et~al.}, arXiv:1205.2360;
%
J.~T.~Hill~{\em et~al.}, arXiv:1206.0704.


\bibitem{OConnell:2010-GroundState}
A.~D. O'Connell {\em et~al.},
\newblock Nature {\bf 464}, 697 (2010).

\bibitem{KolkowitzHarrisLukin:2012-NVc}
S.~Kolkowitz {\em et~al.},
\newblock Science {\bf 335}, 1603 (2012).

\bibitem{OurPhoniton-PRL107}
{\"{O}}.~O. Soykal, R.~Ruskov, and C.~Tahan,
\newblock Phys. Rev. Lett. {\bf 107}, 235502 (2011).

\bibitem{BirPikusBook}
G.~L. Bir and G.~E. Pikus,
\newblock {\em Symmetry and Strain-induced Effects in Semiconductors} (Keter
  Publishing House, Jerusalem, 1974).

\bibitem{Smelyanskiy:2005}
V.~N. Smelyanskiy, A.~G. Petukhov, and V.~V. Osipov,
\newblock Phys. Rev. B {\bf 72}, 081304 (2005).


\bibitem{NewPainterHighQ:2012}
J.~Chan  {\em et~al.},
\newblock Appl. Phys. Lett. {\bf 101}, 081115 (2012).


\bibitem{GoldingDykman:2003}
B.~Golding and M.~I. Dykman,
\newblock arXiv , cond-mat/0309147v1.

\bibitem{Supplementary}
See Supplemental Material.


\bibitem{Luttinger:1956}
J.~M. Luttinger,
\newblock Phys. Rev. {\bf 102}, 1030 (1956).

\bibitem{Neubrand1:1978}
H.~Neubrand,
\newblock Phys. Stat. Solidi (B) {\bf 86}, 269 (1978).

\bibitem{KopfLassmann:1992}
A.~K\"{o}pf and K.~Lassmann,
\newblock Phys. Rev. Lett. {\bf 69}, 1580 (1992).

\bibitem{stress}
Such stress
can be
created in tensioned  
membranes \cite{Teufel:2011-Cooling}. 

%
Larger stress ($\gtrsim 10^7\, \mbox{Pa}$)
%
due to a nearby   (random)
crystal defect \cite{Calvet:2007-LinearStark},
or  SiGe substrate results in a few
hundred GHz splitting.



\bibitem{Calvet:2007-LinearStark}
L.~E. Calvet, R.~G. Wheeler, and M.~A. Reed,
\newblock Phys. Rev. Lett. {\bf 98}, 096805 (2007).

\bibitem{Schuster:2007-PhotonNumberStates}
D.~I. Schuster {\em et~al.},
\newblock Nature {\bf 445}, 515 (2007).


\bibitem{E_Waks:2012}
R.~Bose  {\em et~al.},
\newblock Phys. Rev. Lett. {\bf 108}, 227402 (2012).


\bibitem{Golding:2011}
Y.~P. Song and B.~Golding,
\newblock EPL {\bf 95}, 47004 (2011).


\bibitem{Tezuka:2010-IsotopPurifiedEPR-BSi}
H.~Tezuka  {\em et~al.},
\newblock Phys. Rev. B {\bf 81}, 161203(R) (2010).


\bibitem{StegnerAPL:2011}
A.~R.~Stegner  {\em et~al.},
\newblock Appl. Phys. Lett. {\bf 99}, 032101 (2011).


\bibitem{Karaiskaj:2003-IsotopAcceptorSplit}
D.~Karaiskaj  {\em et~al.},
\newblock Phys. Rev. Lett. {\bf 90}, 016404 (2003).


\bibitem{SafaviNaeini:2011-Translator}
A.~H. Safavi-Naeini and O.~Painter,
\newblock New Journal of Physics {\bf 13}, 013017 (2011).

\bibitem{TianCarmichael:1992}
L.~Tian and H.~J. Carmichael,
\newblock Phys. Rev. B {\bf 46}, R6801 (1992).

\bibitem{MilburnWalls}
D.~F. Walls and G.~J. Milburn,
\newblock {\em Quantum Optics} (Springer, Berlin 2008, 2008).

\bibitem{WallraffTemp}
J.~M. Fink {\em et~al.},
\newblock Phys. Rev. Lett. {\bf 105}, 163601 (2010).


\bibitem{Delsing:2012-SAW}
M.~V. Gustafsson  {\em et~al.},
\newblock Nature Physics {\bf 8}, 338 (2012).


\bibitem{Wallraff:2010-antibunching}
D.~Bozyigit {\em et~al.},
\newblock Nature Physics {\bf 7}, 154 (2011).

\bibitem{DykmanKrivoglaz}
M.~I. Dykman and M.~A. Krivoglaz,
\newblock Sov. Phys. Solid State {\bf 29}, 210 (1987).

\bibitem{Rogge:2004-StarkSi}
G.~D.~J. Smit   {\em et~al.},
\newblock Phys. Rev. B {\bf 70}, 035206 (2004).


\bibitem{SLoth:2010-STM-SpinControlElCurrent}
S.~Loth {\em et~al.},
\newblock Nature Physics {\bf 6}, 340 (2010).

\bibitem{Rabl:2010-indirect}
P.~Rabl {\em et~al.},
\newblock Nature Physics {\bf 6}, 602 (2010).

\bibitem{Morello:2012-SingleAtomSi}
J.~J. Pla {\em et~al.},
\newblock Nature {\bf 489}, 541  (2012).

\bibitem{Braunstein:2003-Transfer}
J.~Zhang, K.~Peng, and S.~L. Braunstein,
\newblock Phys. Rev. A {\bf 68}, 013808 (2003).

\end{thebibliography}
\end{document}